\documentclass[11pt]{article}
\usepackage{blois,epsfig}

\def\Journal#1#2#3#4{{#1} {\bf #2}, #3 (#4)}

\def\NPB{{\em Nucl. Phys.} B}
\def\PLB{{\em Phys. Lett.}  B}

\def\PRD{{\em Phys. Rev.} D}

\def\prep#1#2#3{{\it Phys.\ Rept.} {\bf #1}, #2 (200#3)}
\def\arnps#1#2#3{{\it Ann.\ Rev.\ Nucl.\ Part.\ Sci.} {\bf #1}, #2 (200#3)}

\def\st{\scriptstyle}

\def\ra{\rightarrow}

\def\be{\begin{equation}}
\def\ee{\end{equation}}
\def\bea{\begin{eqnarray}}
\def\eea{\end{eqnarray}}

\newcommand{\Eq}[1]{Eq.~(\ref{eq#1})}
\newcommand{\optbar}[1]{\shortstack{{\tiny (\rule[.4ex]{1em}{.1mm})} 
  \\ [-.7ex] $#1$}}
\def\poptm{\raise.5ex\hbox{${\st +}$}%
    \kern-0.92em\lower1ex\hbox{{\tiny (---)}}} 

\begin{document}
\vspace*{4cm}
\title{TWO QUESTIONS ABOUT NEUTRINOS \footnote{FERMILAB-CONF-10-516-T. To appear in the Proceedings of the 22nd Rencontres de Blois.}}

\author{ BORIS KAYSER }

\address{Theoretical Physics Department, Fermilab, P.O. Box 500, Batavia, IL 60510 USA}

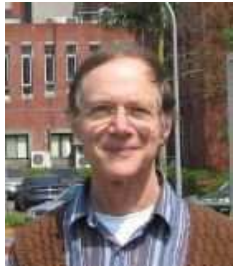
\maketitle\abstracts{
We explain why the see-saw picture and leptogenesis make it particularly interesting to find out whether neutrinos are their own antiparticles, and whether their oscillations violate CP.
}

\section{Motivation}

The see-saw mechanism~\cite{ref1} provides an appealing explanation of the lightness of neutrinos. In its pristine ``type-I'' form, it does this by adding to the Standard Model extremely heavy (beyond LHC range), right-handed, electroweak singlet neutrinos $N$. It then creates an inverse relationship---the see-saw relation---between the masses of these heavy neutrinos and those of the familiar light neutrinos $\nu$.

In the type-I see-saw picture, the only addition to the Standard-Model (SM) Lagrangian is
\be
{\cal L}_{new} = -\frac{1}{2} \overline{{N_R}^c} M_N N_R - (\overline{\nu_L}\overline{\phi^0} - \overline{\ell_L} \phi^- ) y N_R \;+ \mathrm{h.c.} ~~.
\label{eq1}
\ee
One may conveniently assume that the number of heavy neutrinos is three, so that it matches the number of known light SM lepton doublets. Then, in \Eq{1}, $N_R, \nu_L$ and $\ell_L$ are three-component column vectors for the right-handed neutrinos, the SM left-handed neutrinos, and the SM left-handed charged leptons, respectively. The fields $\overline{\phi^0}$ and $\phi^-$ are the usual SM Higgs fields, and $y$ is a 3$\times$3 matrix of Yukawa coupling constants. The first term on the right=hand side of \Eq{1} is a Majorana mass term for the heavy neutrinos. We shall work in a basis in which the Majorana mass matrix $M_N$ that appears in this term is diagonal. The diagonal elements of $M_N$ are then the masses of the heavy neutrino mass eigenstates $N_j$, which must be real numbers, so that $M_N$ must be a real matrix.

The see-saw picture gives rise to a natural explanation of the baryon-antibaryon asymmetry of the universe.~\cite{ref2} Despite their large masses, the heavy neutrinos would have been produced during the hot Big Bang, thanks to its high temperatures. These neutrinos would then have decayed via the Yukawa coupling in \Eq{1}. If the Yukawa coupling matrix $y$ contains CP-violating phases, then in general there would have been a CP-violating difference between the rates for CP-mirror-image decays suh as $N_1 \ra e^- + \phi^+$ and $N_1 \ra e^+ + \phi^-$, where $N_1$ is the lightest of the heavy neutrinos.
As a result, the heavy neutrino decays would have produced a universe with unequal numbers of leptons and antileptons. The non-perturbative SM ``sphaleron'' process, which does not conserve either the net lepton number $L$ or the net baryon number $B$, would then have reprocessed a part of this lepton-antilepton asymmetry into a baryon-antibaryon asymmetry. Detailed analyses have found that this see-saw-inspired scenario, known as leptogenesis, is both qualitatively and quantitatively successful.~\cite{ref3}
It can not only produce a universe with baryons and essentially no antibaryons, but, for very reasonable values of its input parameters, the coupling constants in the Yukawa coupling matrix $y$, it can explain the observed number of baryons per photon.

What experimental information wold add credibility to the see-saw picture or to leptogenesis? Although there are variations, it is a signature feature of the see-saw picture that both the light neutrino mass eigenstates $\nu_j$, and their heavy see-saw partners $N_j$, are their own antiparticles. 
Thus, confirmation that today's light neutrinos are indeed their own antiparticles would increase the plausibility of the see-saw picture, and of leptogenesis, which is a natural outgrowth of that picture. To be sure, leptogenesis cannot occur without CP violation in the leptonic sector. In particular, it requires, as we saw, that the Yukawa coupling matrix $y$ contain CP-violating phases that lead to CP-violating differences between the rates for CP-mirror-image heavy neutrino decays. In the convenient basis in which both the Majorana mass matrix $M_N$ and the charged lepton mass matrix are diagonal and thus real (since particle masses are real), $y$ is the sole source of all CP-violating effects in the leptonic sector. 
If $y$ contains CP-violating phases that give rise to CP violation in heavy neutrino decays, and thereby to leptogenesis, then we expect that in general the phases in $y$ will lead to other leptonic CP-violating effects as well. In particular, as we shall argue, we expect that in general these phases will lead to CP violation in neutrino oscillation. Thus, the observation of CP violation in oscillation would not only establish that CP violation occurs outside the quark sector, but would also make it more plausible that leptogenesis occurred, and that it produced at least a part of the observed cosmic baryon-antibaryon asymmetry.

Clearly, we would like to ask two questions about today's neutrinos: Are they their own antiparticles, and do their oscillations violate CP? To these questions we now turn.

\section{Does $\bar{\nu} = \nu$?}

When Majorana mass terms are present, as they are in the see-saw Lagrangian of \Eq{1}, the neutrino mass eigenstates will be Majorana particles. That is, (for given helicity) each neutrino mass eigenstate will be identical to its antiparticle. One may understand this qualitatively by noting that Majorana mass terms induce neutrino $\leftrightarrow$ antineutrino mixing, and recalling that as a result of $K^0 \leftrightarrow \overline{K^0}$ mixing in the neutral kaon system, the neutral kaon mass eigenstates are not $K^0$ and $\overline{K^0}$, but the states $K_{\mathrm{Short}}$ and $K_{\mathrm{Long}}$. Neglecting CP violation, $K_{\mathrm{Short}} = (K^0 + \overline{K^0} / \sqrt{2}$, and $K_{\mathrm{Long}} = (K^0 - \overline{K^0} / \sqrt{2}$. 
Clearly, apart from an irrelevant sign, each of these states goes into itself under particle $\leftrightarrow$ antiparticle interchange. That is, $K_{\mathrm{Short}}$ and $K_{\mathrm{Long}}$ are identical to their antiparticles. In a similar way, as a result of the $\nu \leftrightarrow \bar{\nu}$ mixing caused by a Majorana mass term, the neutrino mass eigenstate will be of the form $\nu + \bar{\nu}$, which is a state that is identical to its antiparticle.

Since Majorana mass terms mix neutrinos and antineutrinos, they obviously do not conserve the lepton number $L$, which is defined by
\be
L(\nu) = L(\ell^-) = -L(\bar{\nu}) = -L({\ell^+}) = 1 ~~,
\label{eq2}
\ee
so that it distinguishes leptons, neutral and charged, from antileptons. Clearly, if we are to have $\bar{\nu} = \nu, \; L$ cannot be conserved.

Owing to the smallness of the neutrino masses, and to the parity-violating left-handed character of the SM coupling between the neutrinos, the charged leptons, and the $W$ boson, the nature of a Majorana neutrino is somewhat subtle. To clarify this nature, let us first briefly review the nature of a Dirac neutrino, which is a neutrino that is distinct from its antiparticle, and then explain the nature of a Majorana neutrino. We assume our neutrinos are highly relativistic.

A Dirac neutrino mass eigenstate $\nu_j$, plus its antiparticle, is a collection of 4 states with a common mass. With $h$ denoting helicity, these four states are $\nu_j(h=-), \;\bar{\nu}_j(h=+), \;\nu_j(h=+)$, and $\bar{\nu}_j(h=-)$. The left-handed character of the SM weak interaction has the consequence that when a spin-1/2 fermion and its antifermion are distinct and both are highly relativistic, the fermion will interact only when it has left-handed helicity, and the antifermion only when it has right-handed helicity. 
Thus, of the four states that make up a Dirac neutrino and its antiparticle, only the two states $\nu_j(h=-)$ and $\bar{\nu}_j(h=+)$ interact. When $\nu_j(h=-)$ interacts and creates a charged lepton, that lepton will be an $\ell^-$. When $\bar{\nu}_j(h=+)$ interacts and creates a charged lepton, that lepton will be an $\ell^+$. In the case of a Dirac neutrino, we understand this behavior in terms of the lepton number $L$ of \Eq{2}. When neutrinos are of Dirac character, $L$ is conserved and is the property that distinguishes a $\bar{\nu}$ from a $\nu$.

A Majorana neutrino mass eigenstate $\nu_j$ is a collection of just 2 states with a common mass. We may initially call these states $\nu_j(h=-)$ and $\bar{\nu}_j(h=+)$, so that they have the same names as the 2 interacting states in the Dirac case. As in the latter case, when $\nu_j(h=-)$ interacts and creates a charged lepton, that lepton will be an $\ell^-$, but when $\bar{\nu}_j(h=+)$ interacts and creates a charged lepton, it will be an $\ell^+$. However, in the Majorana case, we do {\em not} attribute this behavior to conservation of the lepton number $L$. Indeed, in the Majorana case, $L$ is not conserved, and there is no difference between $\nu_j(h=-)$ and ``$\bar{\nu}_j(h=+)$'' except their helicities. Thus, we will remove the ``bar'' in ``$\bar{\nu}_j(h=+)$'' and call this state simply $\nu_j(h=+)$.
That an interacting Majorana neutrino creates an $\ell^-$ if it has left-handed helicity but an $\ell^+$ if it has right-handed helicity is explained by the parity-violating character of the SM weak interaction. Thanks to this violation of parity, the interactions of the $\nu_j$ states of opposite helicity lead to different final states.

To determine whether neutrino Majorana mass terms occur in nature, so that the neutrinos are Majorana particles, the promising approach is to look for neutrinoless double beta decay ($0\nu\beta\beta$).~\cite{ref4} This is the reaction Nucl $\ra$ Nucl\/$^\prime + e^-e^-$, in which one nucleus decays into another plus two electrons. Clearly, this decay entails $\Delta L = 2$, so that its observation would establish that $L$ is not conserved. In addition, the observation of this decay would establish that nature does contain a Majorana mass term.~\cite{ref5}
To see this, we observe that at the quark level, $0\nu\beta\beta$ is the reaction $dd \ra uu + e^-e^-$. If this reaction is observed, then, by crossing, the amplitude for $e^+ \bar{u} d \ra e^- u \bar{d}$ must be nonzero. 
But, from the SM we know that the amplitudes for the (virtual) reactions $(\bar{\nu})_R \ra e^+ W^-, \; W^- \ra \bar{u}d, \;u\bar{d} \ra W^+$, and $e^-W^+ \ra \nu_L$ are also nonzero. Thus, combining amplitudes, we see that the amplitude for the sequence $(\bar{\nu})_R \ra e^+ W^- \ra e^+ (\bar{u}d) \ra e^- (u\bar{d}) \ra e^-W^+ \ra \nu_L$ must be nonzero. But this sequence adds up to the transition $(\bar{\nu})_R \ra \nu_L$, and this transition is precisely the effect of the Majorana mass term $\overline{\nu_L} {\nu_L}^c$. Consequently, the observation of $0\nu\beta\beta$, at any nonzero level, would establish the existence of an amplitude that is equivalent to a Majorana mass term.

\section{Do Neutrino Oscillations Violate CP?}

In vacuum, the probability $P(\nu_\alpha \ra \nu_\beta)$ that a light neutrino of flavor $\alpha, \; \nu_\alpha$, will oscillate into one of flavor $\beta, \; \nu_\beta$, and the corresponding probability $P(\overline{\nu_\alpha} \ra \overline{\nu_\beta})$ for antineutrinos, are given by
\bea
P(\optbar{\nu_\alpha} \ra \optbar{\nu_\beta}) & =  \delta_{\alpha \beta} \kern-.5em &  - 4\, \sum_{i>j}  \Re\, (U_{\alpha i}^* U_{\beta i} U_{\alpha j} U_{\beta j}^*)\, \sin^2 (\Delta m^2_{ij} \frac{L}{4E})
	\nonumber   \\
& & \poptm 2\,\sum_{i>j} \Im \,(U_{\alpha i}^* U_{\beta i} U_{\alpha j} U_{\beta j}^*)\, \sin (\Delta m^2_{ij} \frac{L}{2E}) ~~.
\label{eq3}
\eea
Here, $\alpha$ and $\beta$ run over the lepton flavors $e,\; \mu$, and $\tau$, $i$ and $j$ run over the light neutrino mass eigenstates, $U$ is the leptonic mixing matrix, $\Delta m^2_{ij} = m^2_i - m^2_j$, where $m_i$ is the mass of mass eigenstate $\nu_i$, $L$ is the distance between the neutrino source and the neutrino detector, and $E$ is the neutrino energy.
From \Eq{3}, we see that if the mixing matrix $U$ is complex, then in general there will be a non-vanishing difference $\Delta_{\alpha \beta} \equiv P(\nu_\alpha \ra \nu_\beta) - P(\overline{\nu_\alpha} \ra \overline{\nu_\beta})$ between the probabilities for corresponding neutrino and antineutrino oscillations. Since $\nu_\alpha \ra \nu_\beta$ and $\overline{\nu_\alpha} \ra \overline{\nu_\beta}$ are CP-mirror-image processes, this non-vanishing $\Delta_{\alpha \beta}$ will be a violation of CP. Its discovery would demonstrate that there is CP violation in the leptonic sector. (Actual experiments will involve neutrino beams traveling through matter, but one may extract from their results the CP-violating vacuum difference  $\Delta_{\alpha \beta}$.)

Assuming the see-saw picture, how is CP violation in neutrino oscillation related to the CP violation in heavy $N$ decay that initiates leptogenesis? The connection between these two CP violations hinges on the fact that the Yukawa coupling matrix $y$ in \Eq{1} plays two roles. First, the coupling constants in $y$ are responsible for the decays of the heavy neutrinos $N$ in the early universe.
Secondly, after the universe cools through the electroweak phase transition and the neutral Higgs field $\phi^0$ develops its nonzero vacuum expectation value $\langle \phi^0\rangle_0 \cong 175 \mathrm{GeV}  \equiv v$, the term $\overline{\nu_L}\overline{\phi^0} y N_R$ in \Eq{1} develops a piece $\overline{\nu_L}(v y) N_R$ which is a ``Dirac mass term'' for the neutrinos. This Dirac mass term is a neutrino analogue of the terms that give masses to the quarks and the charged leptons. The neutrino mass eigenstates, heavy and light, will still be Majorana particles because of the presence of the Majorana mass term in \Eq{1}. When the neutrino sector is diagonalized, the Dirac and Majorana mass terms will combine to yield a diagonal mass matrix $M_\nu$ for the light neutrinos that is inversely related to its counterpart $M_N$ for the heavy neutrinos by the see-saw relation~\cite{ref1}
\bea
M_\nu  &  =  &   - v^2 U^T (y^* M_N^{-1} y^\dagger )\,U       \nonumber  \\
             &  =  &    -v^2 U^T Q  \,U ~~,
\label{eq4}
\eea
where
\be
Q \equiv y^* M_N^{-1} y^\dagger~~.
\label{eq5}
\ee
Here, in our chosen basis, $U$ is the same mixing matrix as the one that appears in the neutrino oscillation probabilities of \Eq{3}, and $M_N$ is the diagonal matrix that appears in \Eq{1}. The diagonal elements of $M_\nu$ and $M_N$ are the masses of the light and heavy neutrino mass eigenstates, respectively. Since particle masses must be real, the matrices $M_\nu$ and $M_N$ must be real.

If leptogenesis is to occur, then, as we have seen, $y$ must contain CP-violating phases. That is, $y$ must be complex. Then the matrix $Q$ defined by \Eq{5} is very likely to be complex as well. But the matrix $M_\nu$ in \Eq{4} must be real. Thus, if the see-saw relation of \Eq{4} is to be satisfied, the mixing matrix $U$ must be complex. From \Eq{3}, we then expect a non-vanishing CP-violating difference between the probabilities $P(\nu_\alpha \ra \nu_\beta)$ and $P(\overline{\nu_\alpha} \ra \overline{\nu_\beta})$. We conclude that if leptogenesis occurred in the early universe, then neutrino oscillation very likely violates CP.

To be sure, it is possible for Q to be real even if $y$ is not real. Then we can have leptogenesis without CP violation in neutrino oscillation. However, for Q to be real when $y$ is not real requires special circumstances that seem rather unlikely, since they would have to somehow relate two pieces of physics that do not appear to be related: the physics of the Majorana mass matrix $M_N$, and the physics of the Yukawa coupling matrix $y$, both of which appear in Q. Thus, we conclude again that if leptogenesis occurred, then neutrino oscillation very likely violates CP.~\cite{ref6}

Clearly, it will be very interesting to find out whether neutrino oscillation does violate CP. Vigorous efforts to design an experimental facility that can achieve this goal are in progress.

Since the other question we have raised is the question of whether $\bar{\nu} = \nu$, one might well ask whether the $\overline{\nu_\alpha} \ra \overline{\nu_\beta}$ and $\nu_\alpha \ra \nu_\beta$ oscillations can differ from one another when $\bar{\nu} = \nu$. The answer is that they can, and that, moreover, the difference between them is independent of whether  $\bar{\nu} = \nu$. To understand this, we note that, in practice, the processes $\nu_\alpha \ra \nu_\beta$ and $\overline{\nu_\alpha} \ra \overline{\nu_\beta}$ are defined, not by the neutrino or antineutrino that travels down the beamline between the source and the detector, but by the particles that couple to this neutrino or antineutrino at the source and the detector. 
For example, to study $\nu_\mu \ra \nu_e$ with conventionally-produced accelerator neutrinos, we would make the neutrino beam via the decays $\pi^+ \ra \mu^+ + \nu_\mu$, and then look for events in which a neutrino in the beam creates an $e^-$ in the detector. To study ``$\overline{\nu_\mu} \ra \overline{\nu_e}$'', we would make the beam via the decays $\pi^- \ra \mu^- +$ ``$\overline{\nu_\mu}$'', and look for events in which a beam particle creates an $e^+$ in the detector. (The detector need not discriminate between $e^-$ and $e^+$. The beam particle from $\pi^+$ decay is left-handed, and consequently cannot create an $e^+$, while the one from $\pi^-$ decay is right-handed, and consequently cannot create an $e^-$.) 
The beam particle is not directly observed, and there is no way of knowing whether, apart from helicity, it is two different beam particles in $\nu_\mu \ra \nu_e$ and  ``$\overline{\nu_\mu} \ra \overline{\nu_e}$'' or not. The amplitudes for these two processes are completely independent of this question, and lead to the same oscillation probabilities, given by \Eq{3}, either way.

In summary, in view of what the see-saw picture and the hypothesis of leptogenesis lead us to expect, it would be very interesting indeed to see whether neutrinos are their own antiparticles, and whether their oscillations violate CP.

\section*{Acknowledgments}

I would like to thank the Scientific Program Committee for having given me the chance to participate in a thoughtfully created and excellent Rencontres de Blois, and to thank Jean Tr\^{a}n Thanh V\^{a}n and Kim Tr\^{a}n Thanh V\^{a}n for exceptional hospitality. Fermilab is operated by the Fermi Research Alliance under contract number DE-AC02-07CH11359 with the U.S. Department of Energy.

\end{document}